%% file: main_Arxiv.tex
\documentclass[11pt,a4paper]{article}

\usepackage[utf8]{inputenc}
\usepackage{authblk} 
\usepackage{geometry} 
\usepackage{graphicx}
\usepackage{amsmath,amssymb,amsfonts}
\usepackage{amsthm}
\usepackage{mathrsfs}
\usepackage[title]{appendix}
\usepackage{xcolor}
\usepackage{textcomp}
\usepackage{manyfoot}
\usepackage{booktabs}
\usepackage{algorithm}
\usepackage{algorithmicx}
\usepackage{algpseudocode}
\usepackage{listings}
\usepackage{braket} 
\usepackage{hyperref} 

\title{Hardware-Free Polarization Stabilization for Measurement-Device-Independent Quantum Key Distribution via Correlated Twirling}

\author[1]{P. Pewkhom\thanks{papon.pe@psu.ac.th}}
\author[2]{N. Jeennugool}
\author[3,4]{N. Ali}
\author[3,4]{R. Endut}
\author[5]{S.A. Aljunid}
\author[1]{P. Kalasuwan\thanks{pruet.k@psu.ac.th}}

\affil[1]{Department of Physical Science, Faculty of Science, Prince of Songkla University, Songkhla, Thailand}
\affil[2]{Triam Udom Suksa School, Bangkok, Thailand}
\affil[3]{Faculty of Electronic Engineering \& Technology, Universiti Malaysia Perlis, Arau, 01000, Perlis, Malaysia}
\affil[4]{Centre of Excellence Advanced Communication Engineering (ACE), Universiti Malaysia Perlis, Arau, 01000, Perlis, Malaysia}
\affil[5]{Faculty of Intelligent Computing, Universiti Malaysia Perlis, Arau, 01000, Perlis, Malaysia}

\date{} 

\begin{document}

\maketitle

\begin{abstract}
Measurement-Device-Independent Quantum Key Distribution (MDI-QKD) provides unconditional security against detector vulnerabilities, but its practical deployment is severely hindered by asymmetric channel turbulence. Fluctuations in optical fibers induce arbitrary polarization drift, degrading Hong-Ou-Mandel interference and forcing extensive calibration downtime. In this work, we propose a hardware-free polarization stabilization technique utilizing a Correlated Twirling protocol based on a unitary 2-design. By applying a synchronized, public twirling supermap, Alice and Bob mathematically transform deterministic, asymmetric geometric rotations into an isotropic Pauli depolarizing channel. Executed entirely as a virtual post-processing step during classical sifting, this protocol mathematically suppresses intrinsic channel noise by a factor of 2/3. We demonstrate through exact quantum state simulations that this induced symmetry neutralizes catastrophic axis-dependent failures, extending the Y-bias tolerance from 0.68 to 0.84 radians. Furthermore, the protocol passively extends the absolute angular misalignment tolerance for the 11\% security threshold from $38.7^\circ$ to $47.9^\circ$, sustaining secure key distillation over extended fiber distances in highly turbulent regimes where standard architectures fail. Inherently compatible with decoy-state weak coherent pulses, this algorithmic approach provides a highly scalable, resource-efficient framework for robust long-distance quantum networks.
\end{abstract}

\vspace{1em}

\input{introduction}
\input{theory_2}
\input{sim_methodology_2}
\input{result_and_discussion_2}
\input{conclusion}

\section*{Acknowledgments}
This work was financially supported by the Faculty of Science Revenue Fund, Prince of Songkla University (Researcher Development Grant 2025, Reference ID: 26344, Project Code: SCI6804154S).

\include{appendix_2}

\bibliographystyle{unsrt} 
\bibliography{bib}

\end{document}

%% file: introduction.tex
\section{Introduction}
\label{sec:introduction}

Quantum Key Distribution (QKD) provides information-theoretic security based on the fundamental laws of quantum mechanics\cite{Portman2022}. However, practical implementations of QKD often deviate from idealized theoretical models, opening the door to side-channel attacks. The majority of these vulnerabilities target the single-photon detectors \cite{Lo2014,Makarov2009,Lydersen2010, Wiechers2011,Sauge2011}. To close this loophole, Measurement-Device-Independent QKD (MDI-QKD) was proposed \cite{Lo2012, Braunstein2012}. In an MDI-QKD architecture, the communicating parties, Alice and Bob, send their prepared quantum states to an untrusted central node (Charlie) who performs a Bell State Measurement (BSM). Because Alice and Bob only rely on their state preparation and Charlie's classical measurement announcements, MDI-QKD is unconditionally secure against all detector side-channel attacks. Connecting all users to the same Charlie is also optimal for a star-type practical quantum network\cite{Valivarthi2017,Zheng2026}. A fully connected multi-user MDI-QKD network has recently been experimentally demonstrated utilizing wavelength multiplexing and integrated optical frequency combs, enabling concurrent safe key exchange across numerous user pairs without the need for trusted nodes\cite{Yan2025}. Numerous experiments have been carried out since the initial proposal of QKD, and the results indicate encouraging progress in three crucial areas: field tests\cite{Tang2016,Tang2015,Cao2020}, key rates\cite{Woodward2021,Hajomer2025}, and channel distance\cite{Tang2014,Yin2016}. Recent advancements demonstrated the capability of integrating classical communication networks with existing fiber optic infrastructure, underscoring the potential for upgrading classical systems to quantum counterparts with MDI-QKD technology\cite{Berrevoets2022}. 

Despite its elegant security advantages and practical implementation, it is essential to highlight the limits of traditional MDI-QKD systems, especially regarding transmission distance and scalability\cite{Zheng2026}. In practical quantum channels, including both optical fiber and free-space links, the transmission efficiency follows an exponential decay with distance, whereas background noise sources, such as detector dark counts and ambient photons, remain nearly constant. Consequently, the signal-to-noise ratio decreases sharply as distance increases, and noise contributions eventually dominate the detection events. This limitation is more severe in MDI-QKD due to its reliance on two-photon interference, which amplifies the impact of channel loss and noise. In terms of secure communication, it also impacts the promotion of Quantum Bit Error Rate (QBER) and a reduction in secret key rates\cite{Peng2025}.

 The efficacy of MDI-QKD, predicated on the BSM, is contingent upon the BSM, which is fundamentally reliant on Hong-Ou-Mandel (HOM) interference at the untrusted node\cite{Hong1987}. For perfect HOM interference to occur, the photons arriving from Alice and Bob must be fundamentally indistinguishable in all degrees of freedom, particularly polarization. In real-world fiber-optic networks, Alice and Bob transmit their photons through independent channels of varying lengths and distinct environmental conditions \cite{Yin2016}. Consequently, the photons experience asymmetric, continuous polarization drifts induced by thermal fluctuations and mechanical stress. 

When the independent unitary rotations of Alice's and Bob's channels diverge ($U_A \neq U_B$), the photons become partially distinguishable. This relative misalignment maps the joint quantum state away from the optimal measurement basis of Charlie's static BSM apparatus, causing a sharp increase in QBER. Beyond a rising QBER, this relative misalignment increases the Guessing Probability ($P_{guess}$), a metric derived from the trace distance between quantum states that quantifies an eavesdropper's ability to distinguish between transmitted signals. As photons become distinguishable, the upper bound of information leakage increases, directly threatening the secret key rate. To overcome this, current implementations rely on three primary paradigms, each with significant drawbacks. Active Polarization Control (APC) systems utilize classical reference pulses and dynamic hardware feedback loops to physically untwist the fibers, but they consume classical resources and introduce severe key-generation downtime \cite{Xavier2011}. Alternatively, Reference-Frame-Independent (RFI) protocols offer a software-based solution by statistically estimating cross-basis errors \cite{Laing2010}. However, RFI operates under the restrictive assumption that the rectilinear ($Z$) basis remains stable, failing under arbitrary 3D mechanical stress. Finally, many architectures abandon polarization entirely in favor of Time-Bin or Phase encoding, which are robust to fiber drift but shift the burden to highly complex, thermally sensitive asymmetric interferometers \cite{Wang2015}.

In this paper, we propose a novel, hardware-free polarization stabilization technique for MDI-QKD that overcomes these limitations by applying a Correlated Measurement Protection protocol. Building upon the concept of optimal measurement preservation via unitary 2-designs , we introduce a physical-layer supermap that revitalizes simple polarization encoding\cite{Ko2025,Kechrimparis2019}. By utilizing a public random beacon to synchronously apply identical twirling operations locally prior to transmission, Alice and Bob effectively transform arbitrary, asymmetric rotational errors into a uniform depolarizing channel through virtual post-processing. This ensures that the protection is implemented purely through classical re-indexing of measurement outcomes, requiring no physical gates after the photons have been sent. This passive approach provides mathematically bounded protection against arbitrary 3D $SU(2)$ rotations without requiring calibration downtime or complex interferometry.

Mathematically, we demonstrate that this correlated twirling crushes the off-diagonal coherence terms of the relative error matrix. The resulting transformation ensures that the surviving fraction of the joint state remains perfectly aligned with the optimal Bell basis:
\begin{equation}
    \mathcal{E}_{eff}(\rho_{rel}) = \frac{1}{|\mathcal{V}|} \sum_{k=1}^{12} V_k^\dagger (U_A^\dagger U_B \rho_{rel} U_B^\dagger U_A) V_k = (1-\eta)\rho_{rel} + \eta \frac{\mathbb{I}}{2}
\end{equation}
where $\mathcal{V}$ represents a 12-element unitary 2-design and $\eta$ is the depolarization parameter dependent on the severity of the turbulence gap. By exchanging a fatal, deterministic rotation for bounded, isotropic white noise, our protocol passively preserves HOM interference. We demonstrate that the protected system exhibits an invariant response to environmental fluctuations. While unprotected systems suffer from volatile performance depending on the specific axis of fiber stress, our protocol ensures the guessing probability follows a stable analytical model: $P_{guess,total,prot} = \left(1 - \frac{2}{3}\sin^2(\alpha/2)\right)^2$. Finally, we present full Monte Carlo simulations of the independent photon evolution, demonstrating that secure MDI-QKD can be sustained under severe, dynamic fiber turbulence.

%% file: theory_2.tex
\section{Theory and Protocol}
\label{sec:theory}

\subsection{The Distinguishability Problem in MDI-QKD}
In a standard MDI-QKD protocol, Alice and Bob independently prepare and transmit single-photon states, denoted as $\rho_A = |\psi_A\rangle\langle\psi_A|$ and $\rho_B = |\psi_B\rangle\langle\psi_B|$, to an untrusted third party, Charlie \cite{Lo2012}. Charlie performs a Bell State Measurement by interfering the incoming photons on a 50:50 beam splitter. 

For the BSM to project correctly onto the Bell basis $\{\Phi^\pm, \Psi^\pm\}$, the photons must undergo perfect Hong-Ou-Mandel interference \cite{Hong1987}. This necessitates that the incident photons are perfectly indistinguishable. However, in practical implementations, the single-mode optical fibers connecting Alice to Charlie and Bob to Charlie induce arbitrary unitary transformations, $U_A$ and $U_B$, respectively, due to thermal fluctuations and mechanical stress \cite{Gisin2002, Xavier2008}. The state arriving at the central node is the joint state:
\begin{equation}
    \rho_{AB}' = (U_A \rho_A U_A^\dagger) \otimes (U_B \rho_B U_B^\dagger)
\end{equation}

The success of the interference is dictated by the relative misalignment between the two channels. We can mathematically express this by analyzing the effective relative rotation operator, $U_{rel} = U_A^\dagger U_B$. When $U_A \neq U_B$, the photons become distinguishable in the polarization degree of freedom. State distinguishability is quantified by the Trace Distance ($T$), which defines the Guessing Probability: $P_{guess} = \frac{1}{2}(1+T)$. 
This continuous, asymmetric rotation shifts the joint state away from Charlie's static BSM measurement axes, introducing off-diagonal coherence terms that manifest as deterministic, elevated mapping errors and an increased capacity for eavesdropper state discrimination.

\subsection{Correlated Twirling via Unitary 2-Designs}
To passively stabilize the measurement basis without active optical compensation, we propose a correlated supermap protocol utilizing a unitary 2-design. A unitary 2-design is a finite set of unitary matrices $\mathcal{V} = \{V_1, V_2, \dots, V_N\}$ that perfectly mimics the mathematical properties of averaging over the entire continuous Haar measure of the $SU(2)$ group for polynomials of degree 2 or less \cite{Gross2007, Dankert2009}. Following the construction in \cite{Ko2025}, we utilize a 12-element set ($N=12$) composed of Pauli matrices and a specific subset of Clifford operations (the explicit operators are detailed in Appendix \ref{app:unitary_operators}).

In the protected protocol, Alice and Bob utilize a synchronized public random beacon to select the exact same unitary $V_k \in \mathcal{V}$ for a given transmission window. The sequence of operations is as follows:
\begin{enumerate}
    \item \textbf{Pre-processing:} Alice and Bob locally apply $V_k$ to scramble their initial states.
    \item \textbf{Transmission:} The scrambled states traverse the independent noisy channels, acquiring the arbitrary rotations $U_A$ and $U_B$.
    \item \textbf{Virtual Post-Processing:} To evaluate the preservation of the measurement basis, we consider the virtual application of the inverse operation $V_k^\dagger$ prior to the Bell State Measurement. 
\end{enumerate}

Consequently, the effective unitary operation applied to Alice's arm is $\tilde{U}_A^{(k)} = V_k^\dagger U_A V_k$, and the effective operation on Bob's arm is $\tilde{U}_B^{(k)} = V_k^\dagger U_B V_k$. 

The probability of successful Hong-Ou-Mandel interference at Charlie's central beam splitter—and thus the success of the BSM—is dictated entirely by the relative misalignment between these two effective operations. By mapping the joint system into a relative reference frame where Alice's effective channel acts as the baseline, the effective relative noise operator $\tilde{U}_{rel}^{(k)}$ acting between the two photons for any given beacon choice $V_k$ can be factored as:
\begin{equation}
    \tilde{U}_{rel}^{(k)} = \left(\tilde{U}_A^{(k)}\right)^\dagger \tilde{U}_B^{(k)} = (V_k^\dagger U_A^\dagger V_k) (V_k^\dagger U_B V_k) = V_k^\dagger (U_A^\dagger U_B) V_k = V_k^\dagger U_{rel} V_k
    \label{eq:relative_u}
\end{equation}

Because the protocol averages over the uniform selection of all 12 unitaries in the 2-design over time, the net effect of the asymmetric channel noise on the distinguishability of the photons is mathematically equivalent to applying a twirled supermap $\mathcal{E}_{rel}$ to the relative quantum state $\rho$:
\begin{equation}
    \mathcal{E}_{rel}(\rho) = \frac{1}{12} \sum_{k=1}^{12} V_k^\dagger U_{rel} V_k \rho V_k^\dagger U_{rel}^\dagger V_k
    \label{eq:twirl_supermap}
\end{equation}

Equation (\ref{eq:twirl_supermap}) is the standard definition of a unitary twirl. By Schur’s Lemma \cite{Nielsen2010} and the defining properties of unitary 2-designs \cite{Dankert2009}, twirling any arbitrary unitary $U_{rel}$ transforms it directly into an isotropic depolarizing channel (the full mathematical derivation of this supermap is provided in Appendix \ref{app:depolarizing_proof}). The protocol transforms the channel into an isotropic depolarizing channel defined by: 
\begin{equation}
    \mathcal{E}_{rel}(\rho) = (1-\eta) \rho + \eta \frac{\mathbb{I}}{2}
    \label{eq:depolarizing_channel}
\end{equation}
where $\mathbb{I}/2$ is the maximally mixed state in the $2 \times 2$ single-photon polarization subspace. Equation (\ref{eq:depolarizing_channel}) illustrates the core protection mechanism: the off-diagonal coherence terms of the relative error are forced to zero, and the fatal geometric rotation is converted into bounded, uniform white noise. Consequently, for the $(1-\eta)$ fraction of surviving photons, perfect relative indistinguishability is restored, and Charlie's fixed BSM apparatus remains the globally optimal measurement basis.

\subsection{Analytical Bounds on Guessing Probability}
\label{sec:analytical_bounds}
To theoretically evaluate the efficacy of the protocol, we analytically derive the Guessing Probability ($P_{guess}$) for both the unprotected and protected systems. The guessing probability is fundamentally linked to the trace distance ($T$) between the target quantum states, defined generally as $P_{guess} = \frac{1}{2}(1 + T)$. Let the relative channel turbulence be parameterized by a misalignment angle $\alpha$ and a spatial rotation axis defined by the unit vector components $n_x, n_y, n_z$. 

For an unprotected MDI-QKD system, the trace distance is heavily dependent on the specific axis of rotation. As detailed in the full formal derivation in Appendix \ref{app:trace_distance}, the individual guessing probabilities for the Bit (Z) and Phase (X) bases are dictated by these rotational components:
\begin{align}
    P_{guess, bit} &= 1 - (1 - n_z^2)\sin^2(\alpha/2) \\
    P_{guess, phase} &= 1 - (1 - n_x^2)\sin^2(\alpha/2)
\end{align}

The total guessing probability in an unprotected system is the product of these two measurements ($P_{guess, total} = P_{guess, bit} \times P_{guess, phase}$). These equations mathematically demonstrate that the unmitigated arbitrary unitary drift creates a theoretically volatile environment; the state distinguishability is not constant, but fluctuates depending on the specific axis of mechanical stress ($n_x, n_z$).

When the Correlated Twirling supermap is applied, the unitary 2-design projects the deterministic rotation into a Pauli depolarizing channel. The depolarization parameter $\eta$ is explicitly defined as $\eta = \frac{4}{3}\sin^2(\alpha/2)$. 

Because the twirling supermap enforces symmetry, the trace distance for the protected system becomes identical across all spatial axes ($T_{prot} = 1-\eta$). Substituting the depolarization parameter yields a stabilized, axis-independent guessing probability for any given basis:
\begin{equation}
    P_{guess,prot\_basis} = 1 - \frac{2}{3}\sin^2\left(\frac{\alpha}{2}\right)
\end{equation}

Since the protected system guarantees equal guessing probabilities across both the bit and phase bases, the rotational axis components ($n_x, n_y, n_z$) are completely eliminated from the state evolution. The total guessing probability is thereby stabilized to a constant envelope regardless of environmental orientation:
\begin{equation}
    P_{guess,total,prot} = \left(1 - \frac{2}{3}\sin^2\left(\frac{\alpha}{2}\right)\right)^2
\end{equation}

By contrasting the respective analytical bounds, it is proven that the twirling supermap inherently transforms a volatile, axis-dependent mapping error into a predictable and strictly bounded metric.

\subsection{Formal Key Sifting and Basis Reconciliation}
\label{sec:key_sifting}

Because the physical twirling operation $V_k$ mathematically scrambles the photon polarization, Charlie's raw Bell State Measurement announcements do not directly correspond to Alice and Bob's original encoded bits. To distill a secure key, the parties must perform a virtual basis reversal during the classical sifting phase. The formal post-processing protocol proceeds in the following sequential steps:

\begin{enumerate}
    \item \textbf{Raw Measurement Announcement:} For each successful coincidence detection, the untrusted node (Charlie) publicly broadcasts the raw measured Bell state $|\psi_C\rangle \in \{\Phi^\pm, \Psi^\pm\}$. If Charlie's detectors fail to register a valid two-photon coincidence, the pulse is discarded.
    
    \item \textbf{Virtual Reversal (The Look-up Phase):} Alice and Bob consult the public random beacon to identify the specific twirling unitary $V_k$ applied during that time bin. Because the joint application of the 2-design unitaries maps the Bell basis strictly onto itself, they classically compute the effective, unrotated state:
    \begin{equation}
        |\psi_{eff}\rangle = (V_k^\dagger \otimes V_k^\dagger) |\psi_C\rangle
    \end{equation}
    This step requires no quantum memory or active optical routing; it is a purely classical deterministic mapping computed via a pre-shared software look-up table (provided in Appendix \ref{app:lookup_table}).
    
    \item \textbf{Basis Sifting:} After the quantum transmission phase is fully completed, Alice and Bob publicly announce their randomly chosen encoding bases ($\beta_A, \beta_B \in \{Z, X\}$) via an authenticated classical channel. They discard all instances where $\beta_A \neq \beta_B$. 
    
    \item \textbf{Bit Extraction and Parity Mapping:} For the remaining instances where their bases match, Alice and Bob extract their raw key bits ($x_A, x_B$). The effective state $|\psi_{eff}\rangle$ dictates the bit correlation (parity) between Alice and Bob:
    \begin{itemize}
        \item If $|\psi_{eff}\rangle \in \{\Phi^+, \Phi^-\}$, the original bits were identical ($x_A = x_B$). Bob leaves his bit unchanged.
        \item If $|\psi_{eff}\rangle \in \{\Psi^+, \Psi^-\}$, the original bits were orthogonal ($x_A = x_B \oplus 1$). Bob applies a classical bit-flip ($X$-gate) to his raw bit to align his key with Alice's.
    \end{itemize}
    
    \item \textbf{Phase Error Estimation:} To bound the information potentially leaked to Eve (or Charlie), Alice and Bob sacrifice a random subset of the data encoded in the $X$-basis to estimate the phase error rate. Because the twirling protocol has transformed the relative rotational drift into an isotropic depolarizing channel, the phase error rate in the $X$-basis acts as a strict, symmetric upper bound on the bit error rate in the $Z$-basis \cite{Shor2000}.
\end{enumerate}

By decoupling the physical noise protection from the measurement apparatus, this sifting protocol ensures that Charlie only ever interacts with the completely randomized states. The cryptographic mapping is strictly isolated within Alice and Bob's classical CPUs, preserving the unconditional security framework of the underlying MDI-QKD architecture.

%% file: sim_methodology_2.tex
\section{Simulation Methodology}
\label{sec:simulation}

To rigorously evaluate the performance of the Correlated Twirling protocol under realistic environmental conditions, we developed a numerical simulation framework using Monte Carlo methods and exact quantum state evolution. The simulations evaluate the protocol across three distinct dimensions: core resilience to channel turbulence, axis-specific phase drift, and distance-dependent attenuation.

\subsection{Quantum State Evolution and The Virtual Framework}
Rather than relying purely on derived analytical bounds, the core engine of the simulation models the exact physical evolution of the quantum states. To maintain proper dimensionality, we define $\rho_{init} = |0\rangle\langle0|$ as the ideal single-photon density matrix prepared by Alice (and identically by Bob). 

To simulate standard, unprotected MDI-QKD, the relative channel noise operator $U_{rel}$ is applied directly to Bob's photon prior to interference. The joint density matrix at the central node is generated by tensoring Alice's unperturbed state with Bob's misaligned state: $\rho_{joint\_unprot} = \rho_{init} \otimes (U_{rel} \rho_{init} U_{rel}^\dagger)$. 

To simulate the Correlated Twirling protocol, the procedure acts as a virtual post-processing step. For each simulated photon pair, a twirling operator $V_k$ is randomly sampled from the predefined 12-element unitary 2-design set $\mathcal{V}$. The effective noise operator acting on Bob's photon is computed as $U_{eff} = V_k^\dagger U_{rel} V_k$, which yields the protected joint state $\rho_{joint\_prot} = \rho_{init} \otimes (U_{eff} \rho_{init} U_{eff}^\dagger)$. The final QBER is extracted by calculating the trace overlap of the respective joint density matrix with the error-flagging Bell state projectors, $P_{\Psi^+}$ and $P_{\Psi^-}$, such that $QBER = \text{Tr}(\rho_{joint} P_{\Psi^+}) + \text{Tr}(\rho_{joint} P_{\Psi^-})$. This method perfectly replicates the deterministic re-indexing performed by Alice and Bob during classical basis sifting.

Simultaneously, the simulation evaluates the fundamental state distinguishability by computing the trace distance ($T$) between the targeted joint states. For each simulated transmission, the trace distance is extracted via $T = \frac{1}{2} \text{Tr}|\rho_{right} - \rho_{wrong}|$, where $\rho_{right}$ and $\rho_{wrong}$ represent the density matrices of the correct and orthogonal measurement projections, respectively. The total guessing probability is subsequently logged as $P_{guess, total} = \frac{1}{2}(1 + T_{bit}) \times \frac{1}{2}(1 + T_{phase})$, serving as the primary metric for information-theoretic stability.

\subsection{Channel Turbulence and Axis-Specific Bias Modeling}
The environmental noise was modeled across three distinct regimes using 2,000 to 5,000 simulated photon pairs per data point to ensure statistical convergence:
\begin{enumerate}
    \item \textbf{Absolute Channel Turbulence:} To evaluate total system failure limits, the relative misalignment angle $\alpha$ was swept linearly from $0^\circ$ to $90^\circ$. The noise operator was defined geometrically as $U_{rel} = \cos(\alpha/2)\mathbb{I} - i\sin(\alpha/2)\sigma_y$.
    \item \textbf{Axis-Specific Bias with Jitter:} To demonstrate the protocol's symmetry, the simulation evaluated discrete rotations specifically along the Polarization (Y) and Phase (Z) axes. The bias was swept from $0$ to $1$ radians. To mimic realistic, high-frequency physical vibrations, a Gaussian noise jitter ($\sigma = 0.02$) was injected into all three spatial rotational components ($n_x, n_y, n_z$) for every individual pulse. 
    \item \textbf{Bounding Environmental Distinguishability:} To evaluate the bounding limits of the guessing probability, the relative rotation angle ($\alpha$) was swept while isolating specific spatial rotational axes ($\mathbf{n}$). The simulation isolated a "Good Axis" (where mechanical stress serendipitously aligns with the measurement basis, maximizing indistinguishability) and a "Bad Axis" (the worst-case orthogonal stress). To simulate a realistic "Turbulent Environment," the spatial axis components ($n_x, n_y, n_z$) were heavily randomized using a uniform Haar-measure distribution for every individual photon pair, representing severe, unpredictable physical fiber vibrations.
\end{enumerate}

\subsection{Distance-Dependent Attenuation and Dark Counts}
To translate the topological mapping errors into practical networking limits, a final simulation evaluated the maximum secure fiber transmission distance. The simulation incorporated standard baseline parameters for Weak Coherent Pulse (WCP) architectures:
\begin{itemize}
    \item \textbf{Fiber Attenuation Coefficient ($\beta$):} 0.2 dB/km (Standard SMF-28 fiber).
    \item \textbf{Mean Photon Number ($\mu$):} 0.5 photons per pulse.
    \item \textbf{Detector Dark Count Probability ($Y_0$):} $10^{-6}$ per detection gate.
\end{itemize}

The total QBER was calculated as the ratio of erroneous detections (arising from intrinsic channel rotation and baseline dark counts) to the total transmission probability. As the fiber distance increases, the exponentially decaying signal probability ($\mu \cdot 10^{-\beta L / 10}$) is eventually dominated by the static dark count probability, pushing the error rate toward $50\%$\cite{Xu2020}. The simulation iteratively searched for the maximum distance ($L_{max}$) at which the total QBER breached the stringent $11\%$ unconditional security threshold dictated by the Shor-Preskill proof, comparing the distance drop-off between raw polarization drift and twirled depolarization.

%% file: result_and_discussion_2.tex
\section{Results and Discussion}
\label{sec:results}

To validate the efficacy of the Correlated Twirling protocol, we analyzed the Monte Carlo simulation data across four critical operational metrics: fundamental state distinguishability, absolute tolerance to angular misalignment, axis-dependent symmetry, and maximum secure transmission distance. The results demonstrate that the application of a unitary 2-design physically stabilizes the quantum channel, closely tracking the analytically derived theoretical bounds.

\subsection{Stabilization of State Distinguishability}
The fundamental root cause of mapping errors in MDI-QKD is the distinguishability of the quantum states arriving at the central node. Before analyzing the bit error rates, we first evaluate the protocol's impact on the total guessing probability ($P_{guess, total}$), which acts as an information-theoretic benchmark for system stability across different environments.

\begin{figure}[htpb]
    \centering
    \includegraphics[width=\linewidth]{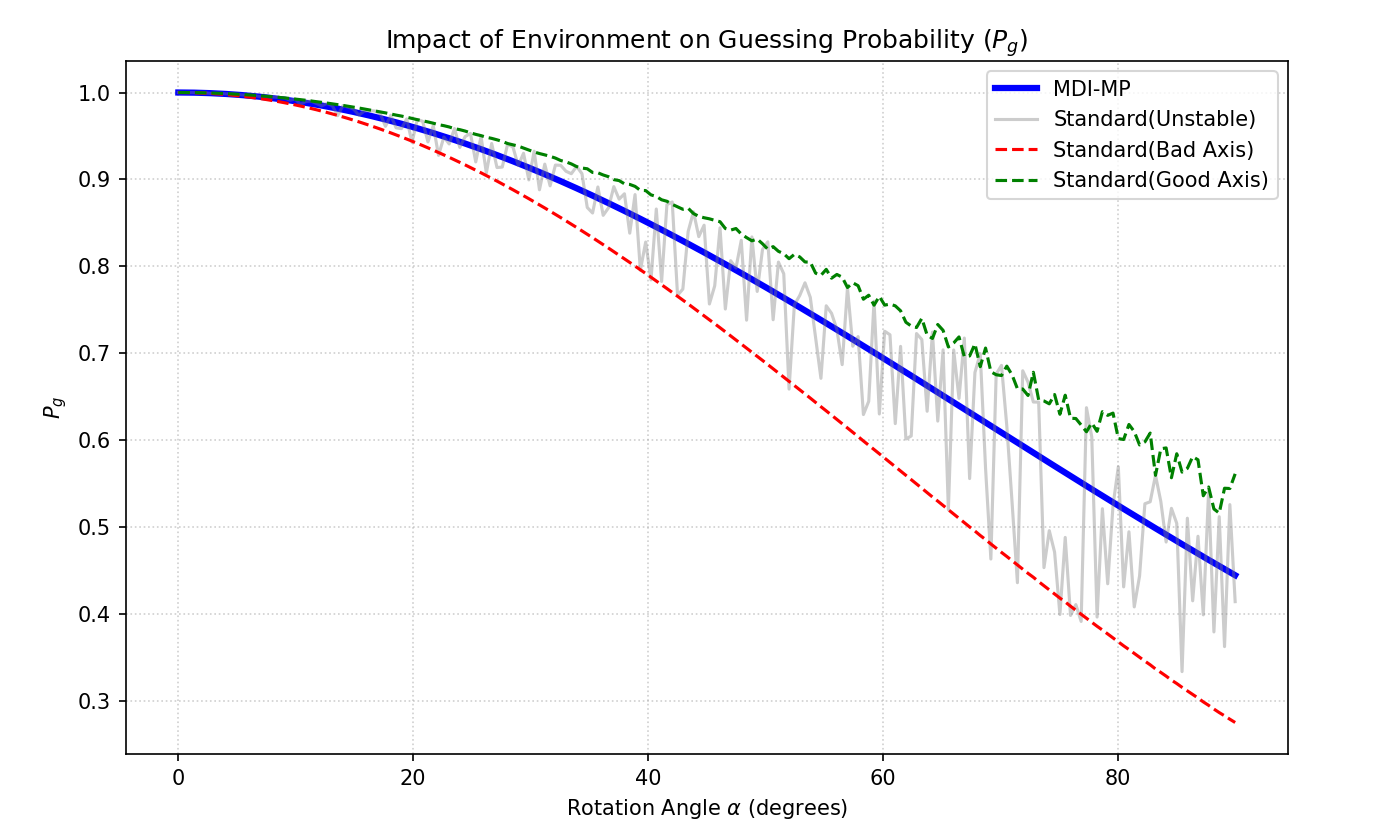}
    \caption{Total guessing probability ($P_{guess, total}$) as a function of the relative rotation angle $\alpha$. The plot compares the invariant response of the protected system (solid blue line) against the volatile performance of the standard unprotected system (gray line), which fluctuates unpredictably between the theoretical worst-case "Bad Axis" (red dashed) and best-case "Good Axis" (green dashed) fiber stress orientations.}
    \label{fig:p_guess_comparison}
\end{figure}

Figure \ref{fig:p_guess_comparison} reveals the severe impact of the spatial rotation axis ($\mathbf{n}$) on unprotected quantum states. In a standard system, the distinguishability is highly unpredictable. If the mechanical stress serendipitously aligns with the measurement basis ("Good Axis"), the system maintains a high guessing probability. However, under orthogonal stress ("Bad Axis"), the distinguishability degrades sharply. In a realistic turbulent environment (represented by the highly volatile gray data), the rotation axis fluctuates randomly, causing the theoretical security bounds to oscillate wildly between these two extremes.

In contrast, the Correlated Twirling protocol exhibits a completely environment-independent response. By transforming the directional geometric error into isotropic depolarizing noise, the protected protocol acts as an active shock absorber. The guessing probability (solid blue curve) is stripped of its axis-dependency, slicing cleanly through the volatility to rigorously follow the derived analytical boundary: $P_{guess, total, prot} = \left( 1 - \frac{2}{3}\sin^2(\alpha/2) \right)^2$. This guaranteed stabilization of the underlying state distinguishability directly suppresses the asymmetric mapping errors observed during the classical sifting phase, which translates to a highly predictable Quantum Bit Error Rate.

\subsection{Resilience to Absolute Channel Turbulence}
\begin{figure}[htpb]
    \centering
    \includegraphics[width=0.85\linewidth]{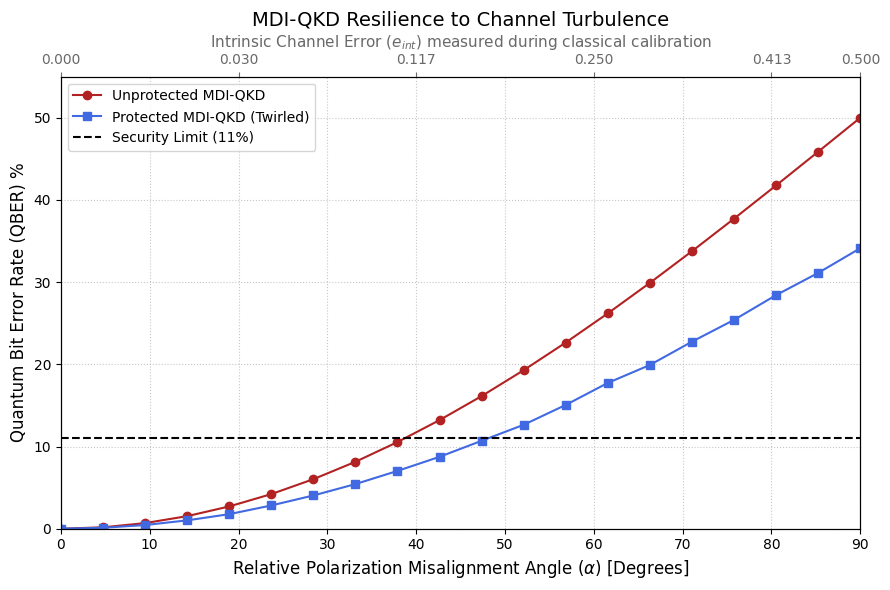}
    \caption{Total Quantum Bit Error Rate (QBER) as a function of the relative polarization misalignment angle ($\alpha$). The unprotected system (red) exhibits severe volatility and breaches the 11\% security threshold at exactly $38.7^\circ$. The application of the Correlated Twirling supermap (blue) suppresses this mapping error into a deterministic envelope, passively extending the operational tolerance to $47.9^\circ$.}
    \label{fig:qber_alpha}
\end{figure}

The primary benchmark for MDI-QKD stability is the system's ability to maintain a Quantum Bit Error Rate (QBER) below the 11\% unconditional security threshold during continuous geometric misalignment. Figure \ref{fig:qber_alpha} illustrates the total QBER as a function of the relative polarization misalignment angle ($\alpha$). 

In the standard, unprotected architecture, the QBER exhibits severe volatility. Because standard MDI-QKD error rates are highly dependent on the specific 3D orientation of the rotational error, the unprotected raw data fluctuates aggressively between zero and maximum intrinsic error. Under worst-case alignment conditions, the unprotected system breaches the 11\% security limit at precisely $\alpha \approx 38.7^\circ$, forcing a total suspension of key generation.

Conversely, the implementation of the Correlated Twirling supermap completely suppresses this volatility. As predicted by the theory, the protected data rigorously follows a deterministic analytical curve. By mathematically forcing the arbitrary geometric error into an isotropic Pauli depolarizing channel, the effective QBER is suppressed by a constant $2/3$ factor. The simulation visually confirms that this virtual protection mechanism acts as an effective shock absorber, smoothly extending the maximum tolerable misalignment angle to $\alpha \approx 47.9^\circ$ without requiring any physical fiber manipulation.

\subsection{Axis-Independent Symmetry and Bias Mitigation}
\begin{figure}[htpb]
    \centering
    \includegraphics[width=\linewidth]{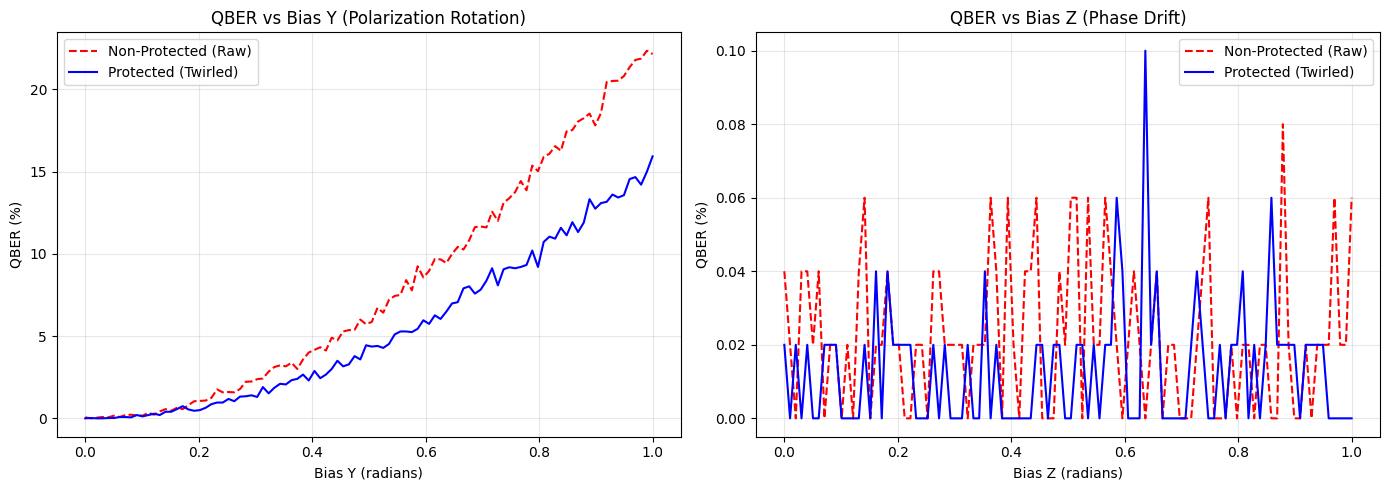}
    \caption{Simulated QBER response to discrete rotational biases applied along the Polarization ($Y$, left) and Phase ($Z$, right) axes. The unprotected system (red, dashed) exhibits severe asymmetrical failure, crossing the 11\% threshold at $0.68$ radians under $Y$-bias, while remaining nearly immune to $Z$-bias. The protected system (blue, solid) enforces error symmetry across all spatial axes, extending the $Y$-bias tolerance to $0.84$ radians.}
    \label{fig:qber_bias}
\end{figure}

The fundamental vulnerability of conventional MDI-QKD lies not only in the magnitude of the error but in its deep asymmetry. Figure \ref{fig:qber_bias} presents the simulated QBER response to discrete rotational biases applied along two orthogonal axes: the polarization bit-flip axis ($Y$) and the phase-drift axis ($Z$). 

Without protection, the system exhibits catastrophic asymmetrical failure. As seen in the left panel of Figure \ref{fig:qber_bias}, standard $Z$-basis MDI-QKD is hyper-sensitive to $Y$-axis rotations, resulting in a sharp, deterministic spike in the raw QBER that violates the 11\% security threshold at exactly $0.68$ radians. However, as shown in the right panel, the exact same magnitude of drift along the $Z$-axis represents pure phase noise, leaving the raw $Z$-basis measurements largely unaffected. This severe topological asymmetry makes parameter estimation and classical tracking loops highly unpredictable in real-world fiber deployments.

The protected simulation data demonstrates the core mathematical power of the unitary 2-design. By virtually applying the twirling set $\mathcal{V}$, the protocol renders the QBER completely invariant to the physical axis of rotation. The protocol willingly sacrifices the artificial "immunity" on the $Z$-axis to mitigate the catastrophic failure on the $Y$-axis. The distinct $Y$ and $Z$ error responses are perfectly symmetrized, transforming arbitrary, multi-axis bias into uniform white noise. Consequently, the protected system smoothly extends the operational tolerance along the critical $Y$-axis to approximately $0.84$ radians. This guaranteed symmetry ensures that the bit error rate strictly mirrors the phase error rate, dramatically simplifying the min-entropy bounds required for privacy amplification.

\subsection{Impact on Secure Transmission Distance}
\begin{figure}[htpb]
    \centering
    \includegraphics[width=0.85\linewidth]{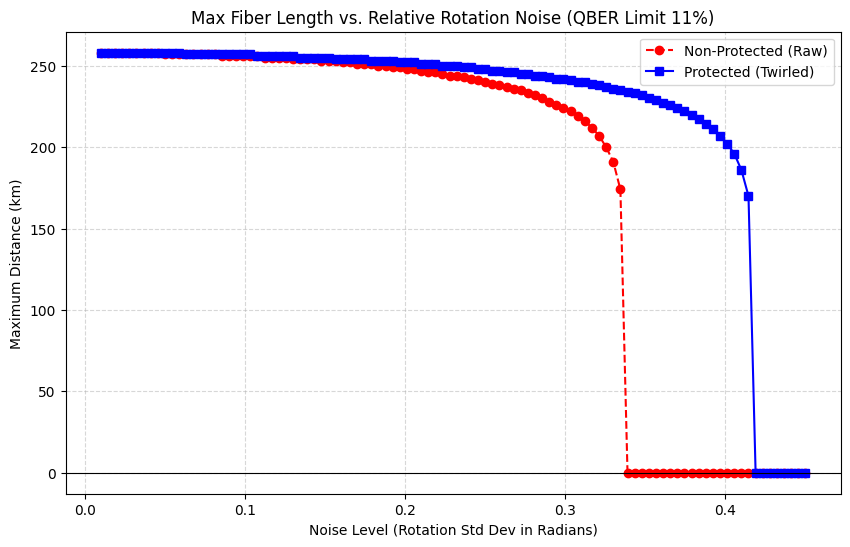}
    \caption{Maximum secure fiber length achievable as a function of relative rotation noise (standard deviation in radians). Both protocols support a baseline transmission distance of 250 km under ideal conditions. However, the unprotected system's operational envelope crashes to 0 km at a noise level of roughly $0.33$ rad. The Correlated Twirling protocol actively suppresses the intrinsic channel noise, sustaining secure key distillation up to a critical failure limit of approximately $0.41$ rad.}
    \label{fig:max_distance}
\end{figure}

While topological error mapping dictates the fundamental physics, network engineers ultimately evaluate QKD protocols by their secure transmission distance. Figure \ref{fig:max_distance} illustrates the maximum secure fiber length achievable under increasing levels of relative rotation noise. 

This simulation dynamically accounts for standard fiber attenuation and baseline detector dark counts. Under highly stable, low-noise conditions, both protocols perform optimally, sustaining a maximum baseline transmission distance of approximately 250 km before the exponential fiber attenuation allows dark counts to dominate the signal-to-noise ratio. 

However, as the environmental noise variance increases, the unprotected system degrades rapidly. The rapid accumulation of geometric errors compounds with the distance-dependent attenuation, causing a precipitous drop in the maximum secure distance. Ultimately, the intrinsic mapping error completely consumes the 11\% security budget before the photons even enter the fiber, driving the secure distance to 0 km at a noise standard deviation of $\sigma \approx 0.33$ radians. 

The Correlated Twirling protocol substantially expands this operational envelope. Because the intrinsic mapping error is mathematically suppressed by the $2/3$ symmetrization factor, a higher fraction of the photon pairs successfully undergo Hong-Ou-Mandel interference at the central node. The protected system reliably resists the escalating turbulence, maintaining a stable transmission distance profile before eventually succumbing to the 11\% threshold at an extended noise limit of $\sigma \approx 0.41$ radians. This confirms that the virtual post-processing protocol not only stabilizes the local measurement basis but directly translates into expanded geographical coverage and reliability for deployable quantum networks.

%% file: conclusion.tex
\section{Conclusion}
\label{sec:conclusion}

In this work, we have proposed and rigorously analyzed a novel, hardware-free polarization stabilization technique for Measurement-Device-Independent Quantum Key Distribution (MDI-QKD). Practical MDI-QKD deployments are fundamentally bottlenecked by the fragility of Hong-Ou-Mandel interference, where asymmetric channel turbulence induces arbitrary polarization drift, elevating the Quantum Bit Error Rate (QBER) and forcing extensive calibration downtime. By leveraging the mathematical properties of unitary 2-designs, our Correlated Twirling protocol offers a passive, algorithmic solution to this physical hardware problem. 

Our theoretical derivations, corroborated by exact quantum state simulations, demonstrate that synchronously applying a public twirling supermap successfully forces arbitrary geometric rotations into an isotropic Pauli depolarizing channel. By crushing the off-diagonal coherence terms of the relative error matrix, the protocol enforces strict error symmetry across all spatial axes. This $2/3$ symmetrization factor yields profound operational advantages: it actively mitigates catastrophic, axis-dependent failures (extending the $Y$-bias tolerance from $0.68$ to $0.84$ radians) and pushes the absolute angular misalignment tolerance of the network from $38.7^\circ$ to $47.9^\circ$. Furthermore, the network simulation confirms that this protocol actively suppresses intrinsic channel noise, sustaining long-distance secure key distillation in turbulent regimes where standard unprotected systems completely fail.

Crucially, because the twirling operators commute with intensity modulation, this protection is inherently compatible with standard decoy-state protocols using Weak Coherent Pulses (WCPs). The required basis reversal is executed entirely as a virtual post-processing step during classical sifting, demanding no quantum memory, dynamic optical routing, or active physical feedback loops. 

By replacing fragile geometric alignment with mathematically guaranteed statistical symmetry, the Correlated Twirling protocol eliminates the downtime associated with active polarization tracking. This provides a highly robust, scalable, and resource-efficient framework for sustaining unconditional security in the next generation of dynamic quantum communication networks. Future work will focus on the experimental validation of this protocol. Initial demonstrations will target short-range proof-of-concept architectures to verify the virtual basis reconciliation, followed by full-scale deployment in long-distance optical fiber networks to evaluate real-time key distillation under dynamic environmental stress.

%% file: appendix_2.tex
\appendix

\section{The Unitary 2-Design Operators}
\label{app:unitary_operators}

The Correlated Twirling protocol utilizes a 12-element unitary 2-design, $\mathcal{V} = \{V_1, V_2, \dots, V_{12}\}$, to optimally average the asymmetric channel noise. This specific 2-design is composed of the Pauli group generators and a symmetrically distributed subset of Clifford rotations. The operators, normalized to preserve unitarity ($V_k V_k^\dagger = \mathbb{I}$), are defined as follows:

{
\renewcommand{\arraystretch}{1.3} 

\textbf{Group 1 (Pauli Operators):}
\begin{align*}
    V_1 &= \begin{pmatrix} 1 & 0 \\ 0 & 1 \end{pmatrix}, &\qquad 
    V_2 &= \begin{pmatrix} 0 & i \\ i & 0 \end{pmatrix}, \\
    V_3 &= \begin{pmatrix} 0 & -1 \\ 1 & 0 \end{pmatrix}, &\qquad 
    V_4 &= \begin{pmatrix} i & 0 \\ 0 & -i \end{pmatrix}
\end{align*}

\textbf{Group 2 (Clifford X-Y-Z Rotations):}
\begin{align*}
    V_5 &= \frac{1}{2}\begin{pmatrix} \phantom{-}1-i & -1-i \\ \phantom{-}1-i & \phantom{-}1+i \end{pmatrix}, &\qquad
    V_6 &= \frac{1}{2}\begin{pmatrix} \phantom{-}1+i & \phantom{-}1-i \\ -1-i & \phantom{-}1-i \end{pmatrix}, \\
    V_7 &= \frac{1}{2}\begin{pmatrix} \phantom{-}1+i & -1+i \\ \phantom{-}1+i & \phantom{-}1-i \end{pmatrix}, &\qquad
    V_8 &= \frac{1}{2}\begin{pmatrix} \phantom{-}1-i & \phantom{-}1+i \\ -1+i & \phantom{-}1+i \end{pmatrix}
\end{align*}

\textbf{Group 3 (Clifford Conjugate Rotations):}
\begin{align*}
    V_9 &= \frac{1}{2}\begin{pmatrix} \phantom{-}1+i & \phantom{-}1+i \\ -1+i & \phantom{-}1-i \end{pmatrix}, &\qquad
    V_{10} &= \frac{1}{2}\begin{pmatrix} \phantom{-}1-i & -1+i \\ \phantom{-}1+i & \phantom{-}1+i \end{pmatrix}, \\
    V_{11} &= \frac{1}{2}\begin{pmatrix} \phantom{-}1-i & \phantom{-}1-i \\ -1-i & \phantom{-}1+i \end{pmatrix}, &\qquad
    V_{12} &= \frac{1}{2}\begin{pmatrix} \phantom{-}1+i & -1-i \\ \phantom{-}1-i & \phantom{-}1-i \end{pmatrix}
\end{align*}
}

\section{Derivation of the Depolarizing Channel via Correlated Twirling}
\label{app:depolarizing_proof}

We mathematically demonstrate that applying the twirled supermap $\mathcal{E}_{rel}$ to the relative noise matrix $U_{rel}$ yields an exact isotropic depolarizing channel. 

First, we parameterize the arbitrary relative unitary operator $U_{rel}$ using the Pauli basis:
\begin{equation}
    U_{rel} = c_0 \mathbb{I} + i \vec{c} \cdot \vec{\sigma}
\end{equation}
where $c_0 = \frac{1}{2}\text{Tr}(U_{rel})$ is a scalar, $\vec{\sigma} = (\sigma_x, \sigma_y, \sigma_z)$ is the Pauli vector, and $\vec{c}$ is a real vector satisfying the unitarity normalization condition $|c_0|^2 + |\vec{c}|^2 = 1$.

Substituting this expansion into the definition of the twirling supermap yields:
\begin{equation}
    \mathcal{E}_{rel}(\rho) = \frac{1}{12} \sum_{k=1}^{12} V_k^\dagger (c_0 \mathbb{I} + i \vec{c} \cdot \vec{\sigma}) V_k \rho V_k^\dagger (c_0^* \mathbb{I} - i \vec{c} \cdot \vec{\sigma}) V_k
\end{equation}

By expanding the multiplication, we obtain three distinct mathematical components: the scalar identity term, the linear cross terms, and the quadratic Pauli terms. 

Because the 12 operators form a unitary 2-design, they perfectly mimic Haar-measure averaging over the Bloch sphere. Consequently, the linear cross terms symmetrically average to zero. The quadratic Pauli term is isotropically randomized, distributing the weight $|\vec{c}|^2$ equally across all three orthogonal Pauli matrices:
\begin{equation}
    \frac{1}{12} \sum_{k=1}^{12} V_k^\dagger (\vec{c} \cdot \vec{\sigma}) V_k \rho V_k^\dagger (\vec{c} \cdot \vec{\sigma}) V_k = \frac{|\vec{c}|^2}{3} \sum_{j \in \{x,y,z\}} \sigma_j \rho \sigma_j
\end{equation}

Applying the single-qubit Pauli identity, $\sum_{j} \sigma_j \rho \sigma_j = 2\mathbb{I} - \rho$, the supermap simplifies to:
\begin{equation}
    \mathcal{E}_{rel}(\rho) = |c_0|^2 \rho + \frac{|\vec{c}|^2}{3} (2\mathbb{I} - \rho)
\end{equation}

Grouping the $\rho$ and $\mathbb{I}$ terms together gives:
\begin{equation}
    \mathcal{E}_{rel}(\rho) = \left( |c_0|^2 - \frac{|\vec{c}|^2}{3} \right) \rho + \frac{4|\vec{c}|^2}{3} \frac{\mathbb{I}}{2}
\end{equation}

We invoke the normalization condition $|\vec{c}|^2 = 1 - |c_0|^2$ to substitute out the vector magnitude in the coefficient for $\rho$, yielding the effective survival fraction $(1-\eta)$:
\begin{equation}
    1-\eta = |c_0|^2 - \frac{1 - |c_0|^2}{3} = \frac{4|c_0|^2 - 1}{3}
\end{equation}

Recalling that $|c_0|^2 = \frac{1}{4}|\text{Tr}(U_{rel})|^2$, we arrive at the exact analytical bound:
\begin{equation}
    1-\eta = \frac{|\text{Tr}(U_{rel})|^2 - 1}{3}
\end{equation}

Therefore, the discrete 12-element twirl mathematically reduces the arbitrary geometric relative rotation into the formal depolarizing channel:
\begin{equation}
    \mathcal{E}_{rel}(\rho) = (1-\eta) \rho + \eta \frac{\mathbb{I}}{2}
\end{equation}
This completes the proof.

\section{Trace Distance Proofs for State Distinguishability}
\label{app:trace_distance}

To evaluate the distinguishability of the quantum states after transmission, we analyze the trace distance ($T$). For two equally likely states $\{\rho_1, \rho_2\}$, the maximum probability of correctly identifying the state (the guessing probability) is given by $P_{guess} = \frac{1}{2}(1 + T)$.

\subsection{Derivation for the Unprotected System}
In an unprotected MDI-QKD system, the joint state is transformed by the relative rotation operator. We parameterize $U_{rel}$ in the Pauli basis as:
\begin{equation}
    U_{rel} = \cos(\alpha/2)\mathbb{I} - i \sin(\alpha/2)(n_x \sigma_x + n_y \sigma_y + n_z \sigma_z)
\end{equation}
This operator can be represented as a matrix $U_{rel} = \begin{pmatrix} a & b \\ -b^* & a^* \end{pmatrix}$, where $a = \cos(\alpha/2) - in_z\sin(\alpha/2)$ and $b = (-n_y - in_x)\sin(\alpha/2)$.

\textbf{Bit Basis (Z-basis):} For the bit basis, we compare the correct joint state $\rho_{right} = \ket{0}\bra{0}_A \otimes U_{rel}\ket{1}\bra{1}U_{rel}^\dagger$ and the incorrect state $\rho_{wrong} = \ket{0}\bra{0}_A \otimes U_{rel}\ket{0}\bra{0}U_{rel}^\dagger$. The trace distance is calculated by the difference in the populations at Bob's side after rotation:
\begin{equation}
    T_{bit} = | |a|^2 - |b|^2 | = |1 - 2(n_x^2 + n_y^2)\sin^2(\alpha/2)|
\end{equation}
Using the normalization $1 = n_x^2 + n_y^2 + n_z^2$, this yields the bit guessing probability:
\begin{equation}
    P_{guess, bit} = 1 - (1 - n_z^2)\sin^2(\alpha/2)
\end{equation}

\textbf{Phase Basis (X-basis):} For the phase basis, we analyze the reduced density matrix difference on Bob's side for the states $\ket{+}$ and $\ket{-}$:
\begin{equation}
    \rho_+ - \rho_- = U_{rel} (\ket{+}\bra{+} - \ket{-}\bra{-}) U_{rel}^\dagger = U_{rel} \sigma_x U_{rel}^\dagger
\end{equation}
The matrix representation of this difference is:
\begin{equation}
    \rho_+ - \rho_- = \begin{pmatrix} 2\text{Re}(ab^*) & a^2 - b^2 \\ (a^*)^2 - (b^*)^2 & -2\text{Re}(ab^*) \end{pmatrix}
\end{equation}
The trace distance is the absolute difference in the diagonal terms when projected onto the $\ket{+}$ basis:
\begin{equation}
    T_{phase} = |\text{Re}(a^2 - b^2)| = \left| \cos^2(\alpha/2) - n_z^2\sin^2(\alpha/2) - (n_y^2 - n_x^2)\sin^2(\alpha/2) \right|
\end{equation}
Applying trigonometric identities and spatial normalization yields:
\begin{equation}
    T_{phase} = |1 - 2\sin^2(\alpha/2)(n_y^2 + n_z^2)|
\end{equation}
This results in the phase guessing probability:
\begin{equation}
    P_{guess, phase} = 1 - (1 - n_x^2)\sin^2(\alpha/2)
\end{equation}

\subsection{Derivation for the Protected System}

In the protected protocol, the correlated twirling operation transforms the directional rotation into an isotropic depolarizing channel defined by $\mathcal{E}_{rel}(\rho) = (1-\eta)\rho + \eta\frac{\mathbb{I}}{2}$, where the depolarization parameter is strictly $\eta = \frac{4}{3}\sin^2(\alpha/2)$.

To find the distinguishability, we compare the evolved states:
\begin{align}
    \rho_{right} &= \ket{0}\bra{0}_A \otimes \mathcal{E}_{rel}(\ket{1}\bra{1}) = \ket{0}\bra{0}_A \otimes \left((1-\eta)\ket{1}\bra{1} + \eta\frac{\mathbb{I}}{2}\right) \\
    \rho_{wrong} &= \ket{0}\bra{0}_A \otimes \mathcal{E}_{rel}(\ket{0}\bra{0}) = \ket{0}\bra{0}_A \otimes \left((1-\eta)\ket{0}\bra{0} + \eta\frac{\mathbb{I}}{2}\right)
\end{align}
The distinguishability is the trace distance between these joint states:
\begin{equation}
    T_{prot} = \frac{1}{2} \text{Tr} |\rho_{right} - \rho_{wrong}| = 1-\eta
\end{equation}
Substituting the analytical value of $\eta$ yields:
\begin{equation}
    T_{prot} = 1 - \frac{4}{3}\sin^2(\alpha/2)
\end{equation}
Because the initial trace distance $T$ for both the Z-basis and the X-basis is exactly 1, the resulting trace distance after twirling protection is identically $1-\eta$ for all bases. Consequently, the guessing probabilities are equal ($P_{guess, bit} = P_{guess, phase}$), leading to an axis-independent guessing probability of:
\begin{equation}
    P_{guess, prot\_basis} = \frac{1}{2}(1 + T_{prot}) = 1 - \frac{2}{3}\sin^2(\alpha/2)
\end{equation}

\section{Virtual Post-Processing Look-Up Table}
\label{app:lookup_table}

During the classical sifting phase, Alice and Bob virtually apply the inverse operation $V_k^\dagger \otimes V_k^\dagger$ to deduce the unrotated Bell state from Charlie's raw measurement announcement. Table \ref{tab:lookup} details the deterministic mapping used to perfectly reverse the physical scrambling without active optical components.

\begin{table}[htpb]
    \centering
    \caption{Classical Look-Up Table for Correlated Twirling Post-Processing}
    \renewcommand{\arraystretch}{1.3}
    \begin{tabular}{l | c c c c}
        \hline\hline
        \textbf{Public Beacon} & \textbf{If True $\Phi^+$} & \textbf{If True $\Phi^-$} & \textbf{If True $\Psi^+$} & \textbf{If True $\Psi^-$} \\
        \hline
        $V_1$ to $V_4$ (Pauli) & $\Phi^+$ & $\Phi^-$ & $\Psi^+$ & $\Psi^-$ \\
        $V_5$ to $V_8$ (Clifford) & $\Phi^+$ & $\Psi^+$ & $\Phi^-$ & $\Psi^-$ \\
        $V_9$ to $V_{12}$ (Clifford) & $\Phi^-$ & $\Phi^+$ & $\Psi^-$ & $\Psi^+$ \\
        \hline\hline
    \end{tabular}
    \label{tab:lookup}
\end{table}